\documentclass[aps,amsmath,twocolumn,amssymb,floatfixng,showpacs,
superscriptaddress,footinbib]{revtex4-1}
\usepackage[T1]{fontenc}
\usepackage[utf8]{inputenc}

\pdfoutput=1
\usepackage[dvips]{graphics}
\usepackage{bm}
\usepackage{float}
\usepackage{epsfig}
\usepackage{enumerate}
\usepackage{subfigure}
\usepackage{amsmath}
\usepackage{color}
\usepackage{braket}
\usepackage{graphicx}
\usepackage{float}
\usepackage[colorlinks=true,linktoc=page,linkcolor=red,citecolor=blue,urlcolor=magenta]{hyperref}

\newcommand{\myvec}[1]{\textbf{#1}}

\newcommand\bea{\begin{eqnarray}}
\newcommand\eea{\end{eqnarray}}
\newcommand\beq{\begin{equation}}  
\newcommand\eeq{\end{equation}}

\begin{document}
\title{Eight fold  quantum Hall phases in a time reversal symmetry broken
tight binding model}
\author{Sudarshan Saha}\email{sudarshan@iopb.res.in}\affiliation{Institute of Physics, Bhubaneswar- 751005, Odhisa, India}
\affiliation{Homi Bhabha National Institute, Mumbai - 400 094, Maharashtra, India}
\author{Tanay Nag}\email{tnag@physik.rwth-aachen.de}\affiliation{Institute f\"ur Theorie der Statistischen Physik, RWTH Aachen University, 52056 Aachen, Germany}
\author{Saptarshi Mandal}\email{saptarshi@iopb.res.in}\affiliation{Institute of Physics, Bhubaneswar- 751005, Odhisa, India}
\affiliation{Homi Bhabha National Institute, Mumbai - 400 094, Maharashtra, India}
	
\begin{abstract}

We consider a time reversal symmetry (TRS) broken Kane-Mele model superimposed with Haldane model 
and chart out the phase diagram using spin Chern number to  investigate the fate of quantum anomalous Hall insulator (QAHI) and quantum  spin Hall insulator (QSHI) phases. Interestingly, in addition to QSHI and QAHI phase, the phase diagram unveils  quantum anomalous spin Hall insulator (QASHI) phase where only one spin sector is topological. We also find  multicritical points where  three / four topological phase boundaries coalesce.  These topological phases are protected by an effective TRS and a composite anti-unitary particle-hole symmetry leading to remarkable properties of edge modes. We find spin-selective, spin-polarized and spin-neutral edge transport in QASHI, QSHI and QAHI phases respectively. Our study indicates that the robustness of the topological phase mainly depends on the spin gap which does not necessarily  vanish at the Dirac points across a topological phase transition. We believe that our proposals can be tested in near future using recent experimental advancements in solid state and cold atomic systems.
\end{abstract}

	
\date{\today}
	
\maketitle

  Recently various  non-interacting quantum Hall systems such as, quantum anomalous Hall insulator (QAHI) \cite{haldane-1988,nagaosa-2003} and quantum spin Hall insulator (QSHI) \cite{kane-2005-1st,kane-2005-2nd,barnevig-2006} and many more \cite{QHE_review1,QHE_review2,QHE_review3,QHE_review4} have been investigated in various topological context. The QAHI and QSHI are best characterized by the quantized charge and spin current respectively. This is intimately connected with the fact that QAHI \cite{haldane-1988} breaks  TRS while QSHI  does not \cite{kane-2005-1st}. 
  The spin-orbit coupling serves as a basic ingredient for QSH effect to occur.  It has been shown that the bulk topological invariant Chern number \cite{niu-85} (spin Chern number \cite{haldane06}) can successfully predict the number of edge states (spin polarized channels) in QAHI (QSHI) phases \cite{bbc,li12}. Another widely used topological invariant namely, $Z_2$ index can equivalently  classify TR invariant system \cite{kane-2005-1st,prodan-2009,spin-chern1,spin-chern2,spin-chern3,spin-chern4}. 
 Thereafter it becomes an important question that what would be the fate of the QSHI phase in the absence of TRS.

 In order to search for the  answers, TRS breaking terms such as, exchange field \cite{qiao10,sheng-2011}, magnetic doping \cite{Liu-2008,li2013}, and staggered magnetic flux \cite{luo-2017} are introduced in QSH system to obtain  QAH effect. Remarkably, even though $Z_2$ index fails to characterize the topological nature of the phase, spin Chern number persists to  be a relevant topological invariant distinguishing a TRS broken QSHI phase from a QAHI phase. 
The QSH \cite{kane-2005-1st}  and QAH \cite{haldane-1988} models have been generalized to various theoretical platforms \cite{kim16,rachel14} and realized in experiments \cite{exp1,exp2}. 
All these studies 
motivate us to consider Kane-Mele model 
infused with Haldane model such that the TRS is broken by  staggered magnetic flux associated with next nearest neighbour (NNN) hopping and intrinsic SOC term.  To be precise, we ask the  following questions: 1. How do the Haldane and Kane-Mele phase diagrams modify? 2. Are there any new topological phases apart from QSHI and QAHI phases? 3. Can spin Chern number successfully describe all the phases?

\begin{figure}[!htb]

\includegraphics[width=0.23\textwidth]{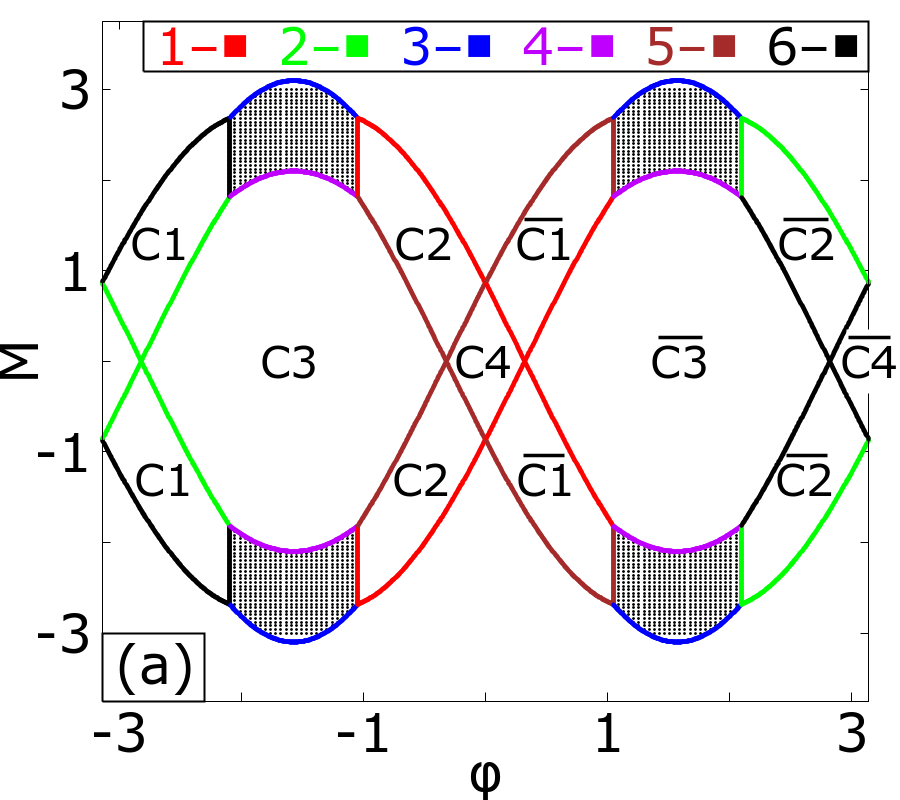}
\includegraphics[width=0.23\textwidth]{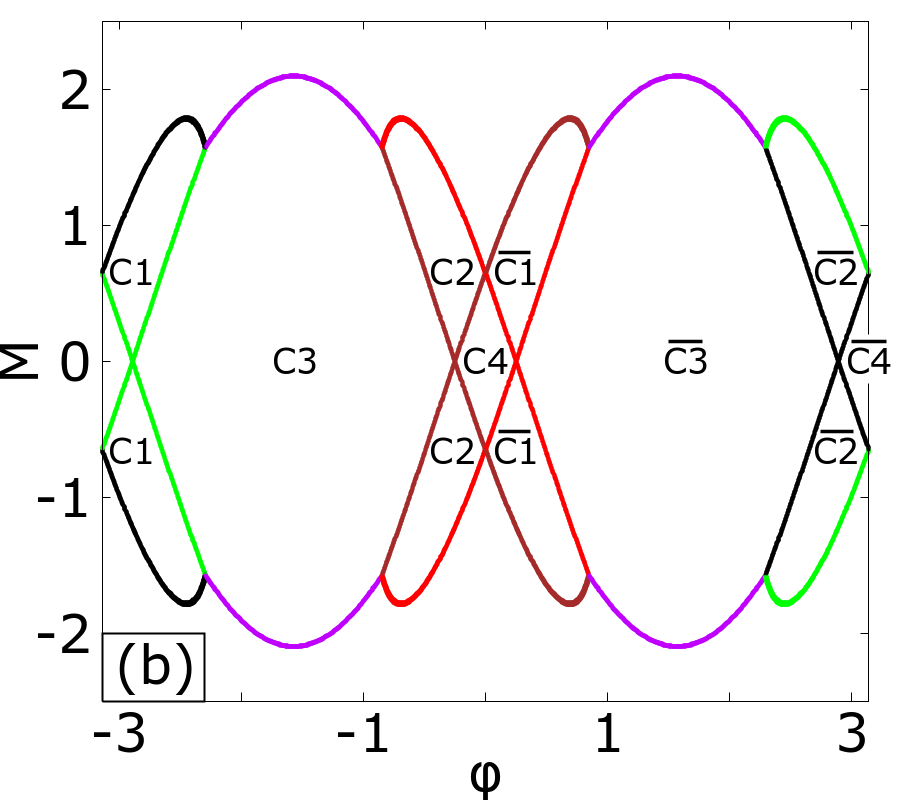}

\caption{Here we  show the effect of $V_R$ for a fixed $V_{\rm so}=1.0$ in $M$-$\phi$ phase diagram. (a) and (b) are plotted with $V_R=0.0$ and $V_R=0.5$ respectively. The indices within a given phase refer the values of $(\rm{C}_{\uparrow}, \rm{C}_{\downarrow}$): $\rm{C}1=(0,1)$, $\rm{C}2=(1,0)$, $\rm{C}3=(1,1)$, $\rm{C}4=(1,-1)$ and $\overline {\rm{C}n}=-\rm{C}n$ with $n=1,2,3,4$.
The phase boundary is obtained by the zeros of band gap \cite{phase_boundary} and the color codes refer to the relevant band gap equations. 
The spin gap vanishes in critical phase as denoted by the assembly of black dots in (a).}
\label{fig:modificationHaldane}
\end{figure}

\begin{figure}[!htb]

 \includegraphics[width=0.47\linewidth]{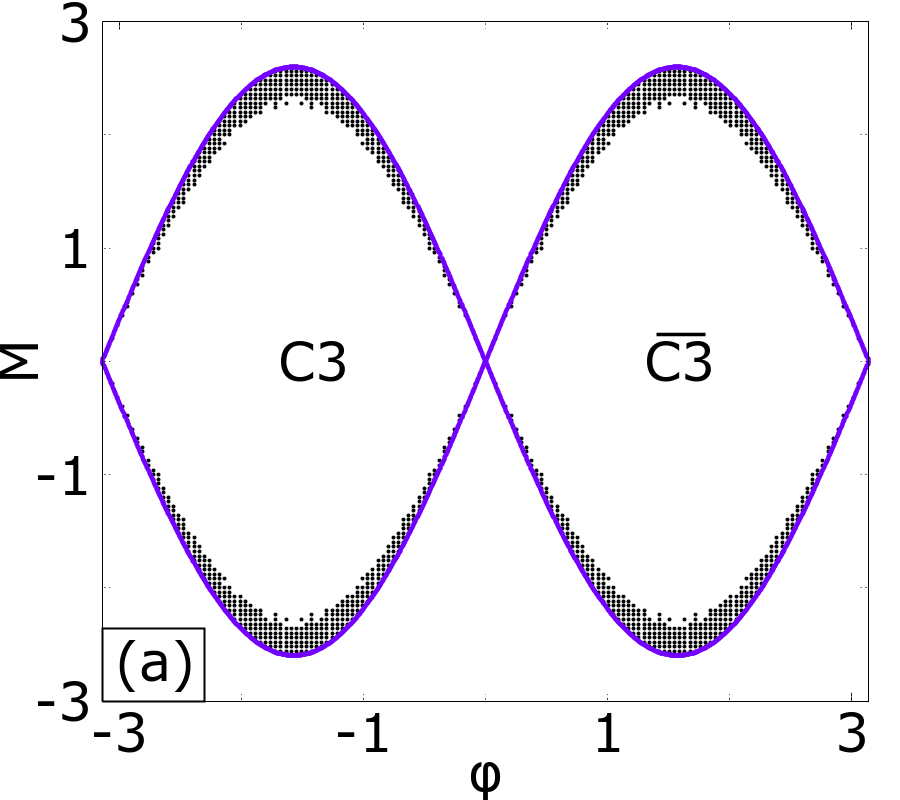}
 \includegraphics[width=0.47\linewidth]{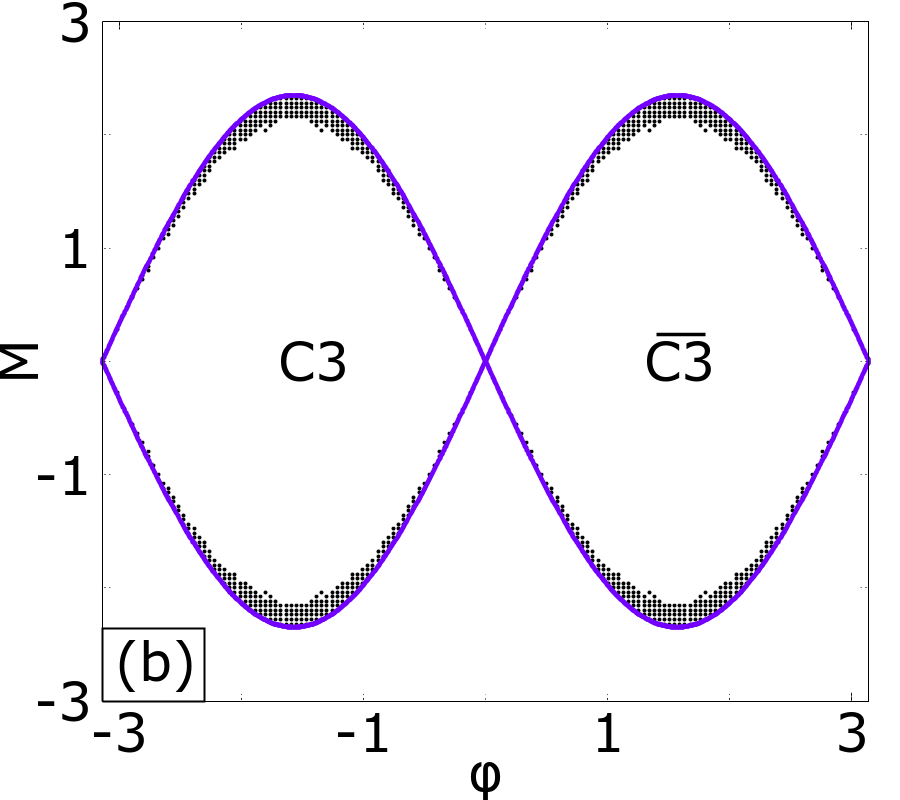}

\caption{Here we investigate the effect of $V_{\rm so}$ for a fixed $V_R=1.0$ in $M$-$\phi$ phase diagram. (a) and  (b) are plotted for $V_{\rm so}=0.0$ and $V_{\rm so}=0.5$ respectively. The definition of $\rm{C}n$ and $\overline {\rm{C}n}$ are provided in the caption of Fig.~\ref{fig:modificationHaldane}.
The spin gap vanishes in critical phase, denoted by the assembly of black dots, that gets narrower with increasing $V_{\rm so}$. }
\label{fig:modificationHaldane2}
\end{figure}

\begin{figure}[!htb]

 \includegraphics[width=0.48\linewidth]{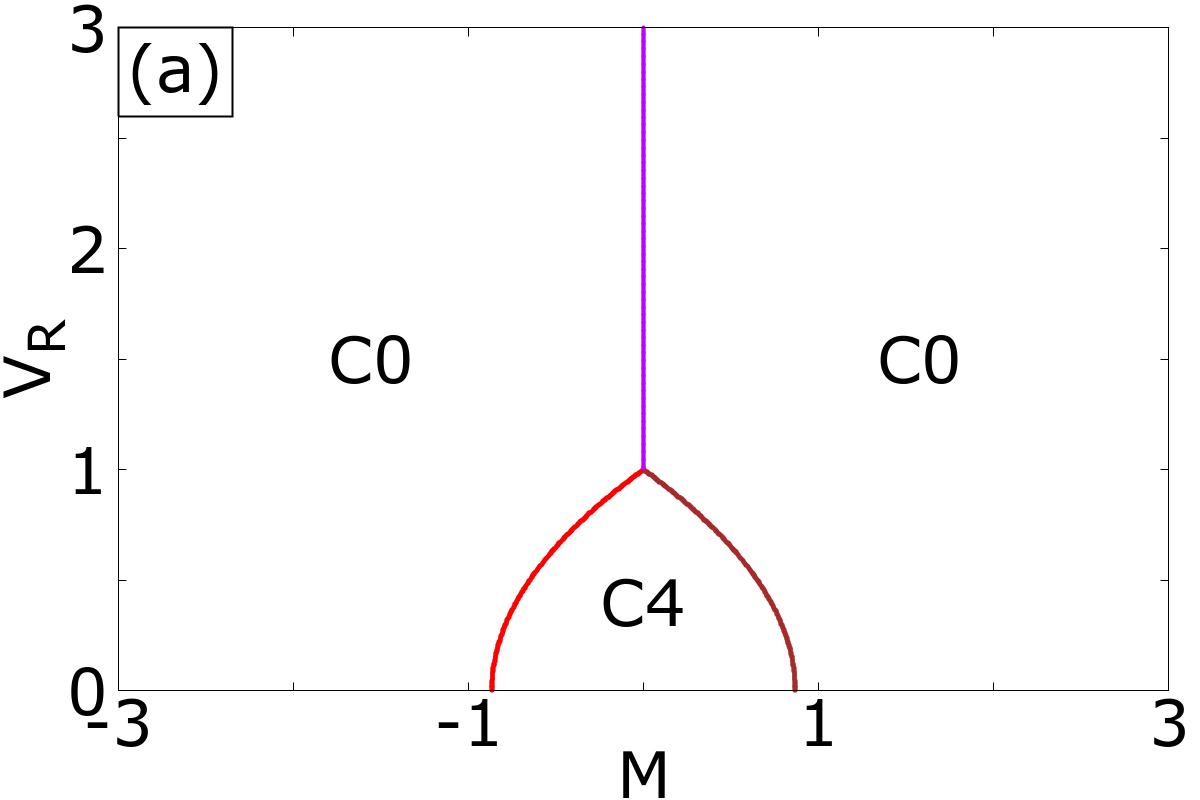}
 \includegraphics[width=0.48\linewidth]{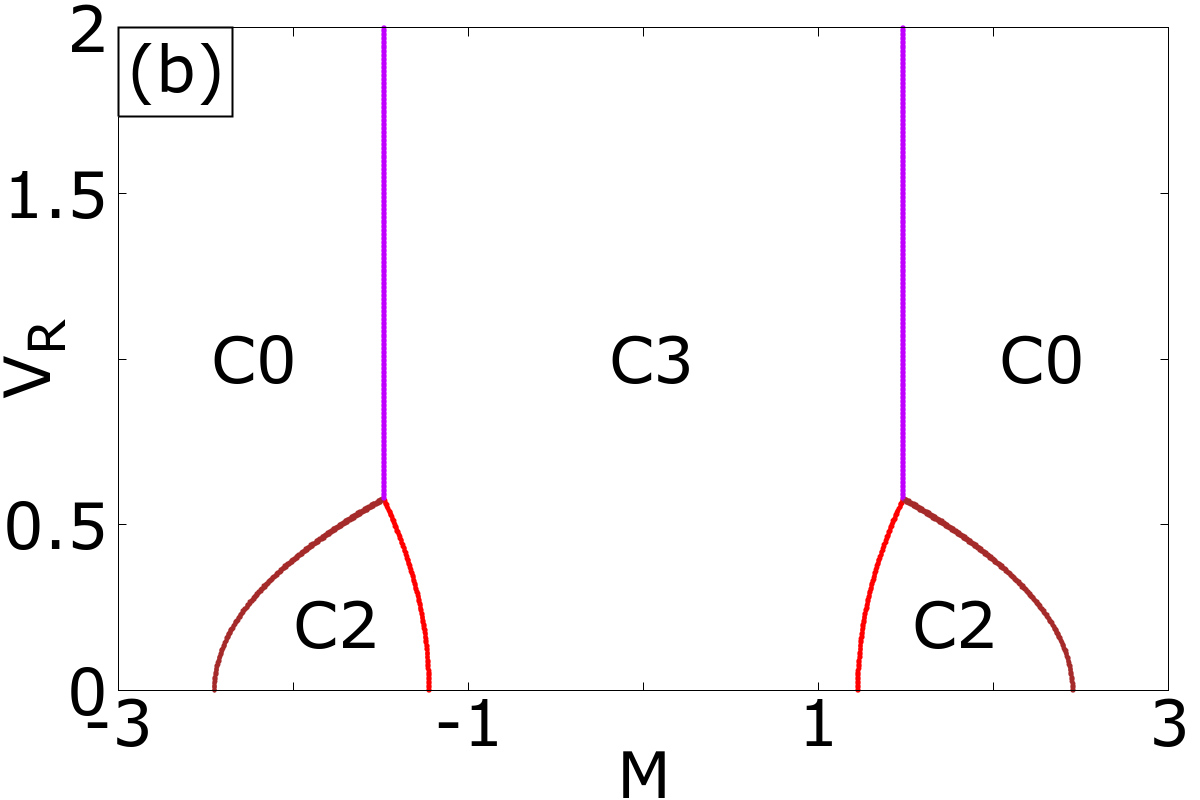}

\caption{ We here demonstrate that how a QSHI phase gives rise to  QAHI and QASHI phase by varying $\phi$ in $M$- $V_{R}$ plane keeping $V_{\rm so}=1.0$ fixed. (a) and (b)correspond to $\phi=0.0$ and $\phi= -\pi/4$, respectively.
The definition of $\rm{C}n$ and $\overline {\rm{C}n}$ are provided in the caption of Fig.~\ref{fig:modificationHaldane}.}
\label{fig:modificationKaneMele}
\end{figure}

\begin{figure}[!htb]

 \includegraphics[width=.95\linewidth]{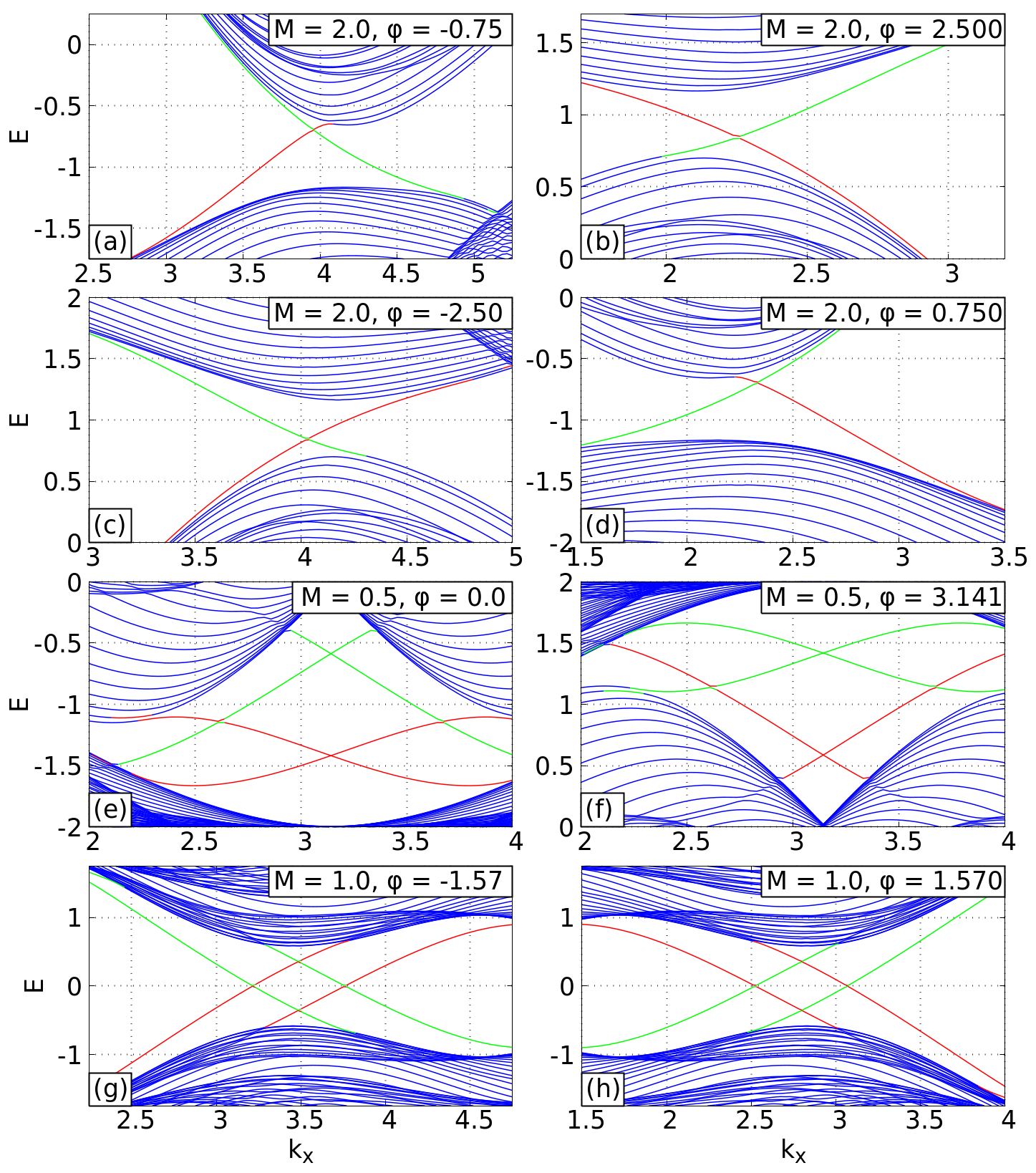}
 
\caption{ Here we display the edge modes for various topological phases: QASHI phase in (a), (b), (c), (d) for $(1,0)$, $(-1,0)$,
$(0,1)$ and  $(0,-1)$, respectively; QSHI phases in (e) and (f) for  $(1,-1)$ and $(-1,1)$, respectively; QAHI phases in (g) and (h) for $(1,1)$ and $(-1,-1)$, respectively.  The red (green) refers to the localization of edge modes at top (bottom) part of the semi-infinite zig-zag chain. 
}
\label{fig:edgeStateVsoGTVr}
\end{figure}

To this end, we first demonstrate  how the Haldane (Kane-Mele) phases evolve with the  Rashba and SOC terms (NNN hopping and magnetic flux) [see Fig.~\ref{fig:modificationHaldane}, Fig.~\ref{fig:modificationHaldane2} and Fig.~\ref{fig:modificationKaneMele}]. The zeros of bulk energy gap determine the topological phase boundaries and the  finite spin gap \cite{sheng-2011,spin-gap} provides   the  robustness of spin Chern number $(\rm{C}_{\uparrow},\rm{C}_{\downarrow})$ \cite{prodan-2009,spin-chern-number} of a topological phase. 
The QASHI [QSHI] phases are denoted by $(\rm{C}_{\uparrow},0)$ and $(0,\rm{C}_{\downarrow})$ [$(\rm{C}_{\uparrow},\rm{C}_{\downarrow})$ with $\rm{C}_{\uparrow} \ne \rm{C}_{\downarrow} \ne 0$] while QAHI phase is designated by $\rm{C}_{\uparrow}=\rm{C}_{\downarrow} \ne 0$. 
In confirmation of bulk-boundary correspondence, we find spin-selective, -polarized and -neutral transport in QASHI, QSHI and QAHI phases in the band structure in semi-infinite geometry with zig-zag edge (see Fig.~\ref{fig:edgeStateVsoGTVr}). These findings  are further explained by the low energy version of the model where  evolution of spin dependent Haldane gap  with various parameters are demonstrated.    
These topological phases are protected under emerging anti-unitary symmetries that couple with the chirality of the flux. In essence, considering a simple flux induced  TRS broken QSHI
model, our study uncovers many extraordinary features in a systematic manner for the first time and also  opens up the possibility of practical device applications in future.

The Hamiltonian we consider here is given below,
\begin{eqnarray}
\label{main:ham}
H = && -t_1 \sum_{\langle ij \rangle} c^{\dagger}_i c_j + iV_R \sum_{\langle ij \rangle} c^{\dagger}_i (\vec{\sigma} \times \vec{d}_{ij})_z c_j + M \sum_i c^{\dagger}_i \sigma_z c_i  \nonumber \\
&&~ + t_2 \sum_{\langle\langle ij \rangle\rangle} e^{i \phi_{ij}} c^{\dagger}_i c_{j} + \frac{i V_{\rm so}}{\sqrt{3}} \sum_{\langle\langle ij\rangle\rangle}e^{i \phi_{ij}} \nu_{ij} c^{\dagger}_i \sigma^z c_j 
\end{eqnarray}
where $c_i$ represents the fermion spinor $(c_{i\uparrow}, c_{i\downarrow})$; $V_{\rm so}$ and $V_R$ represent the SOC and Rashba interaction strength, respectively. The model incorporates a spin-independent NN (NNN) hopping denoted by $t_1$ ($t_2$). The phase factor $e^{i \phi_{ij}}$ comes due to the staggered magnetic flux as described  in the Haldane model \cite{haldane-1988}. The factor $\nu_{ij}=(\myvec{d}_{ij}^1\times\myvec{d}_{ij}^2)_z $ and the lattice vectors are same as mentioned in Kane-Mele model ~\cite{kane-2005-2nd}. The important point to note here is that the SOC term acts as the spin dependent NNN hopping of strength $V_{\rm so}$ where 
spin-dependent magnetic fluxes are essentially coupled to the electron momenta.  $M$ represents the inversion breaking mass term. \\
\indent

{Before proceeding further, we emphasize the key features of the model Hamiltonian as given in Eq. (\ref{main:ham}) and  their physical implications in detail.  We know that in the limit $V_R=V_{\rm{so}}=0$, the model contains two copies of Haldane model.  In this case the spin indices are irrelevant. The topological phases follow  the condition $M < | 3 \sqrt{3} t_2 \sin \phi|$ \cite{haldane-1988}. The TRS is broken there and both the spin sectors have the same spin Chern number. To understand the phases, obtained in Kane-Mele model, we note that they correspond to $\phi=0, t_2=0$, but do contain the next nearest neighbour spin dependent hopping with complex magnitudes. The fact that the complex NNN spin dependent hopping changes sign under spin flipping implying the restoration of the TRS. 
The spin Chern numbers of opposite spin sectors are opposite.  The total spin Chern number must add up to zero owing to TRS invariant nature  of the system. As long as TRS is preserved additional interaction such as Rashba spin-orbit interaction is not able to change the scenario and new phases will not appear. Interestingly when the TRS is broken by introducing a flux in the same spirit of Haldane model, one expects new topological phases to appear. In particular, we introduce the magnetic flux in the SOC term so that the spin dependent NNN hopping acquires complex amplitudes.
The sum of spin Chern number over all the  spin sectors is no longer constrained to be zero. This observation opens up the possibility of QASHI phase where one spin sector is topological and other is not.}

We now discuss in detail the phase diagram obtained by investigating the Hamiltonian (\ref{main:ham}) in momentum space.  One can obtain the momentum space Hamiltonian after Fourier transformation of Eq.~(\ref{main:ham}) as given by,
\begin{equation}
H({\bm k})= \sum_{i=0}^9 n_i (\bm k) ~ \Gamma_i 
\label{main:hamk}
\end{equation}
with $\Gamma_i=\sigma_{i} \otimes \tau_0$ for $i=1,2,3$, $\Gamma_{i+3}=\sigma_{i} \otimes \tau_1$ for $i=1,2$, $\Gamma_{i+5}=\sigma_{i} \otimes \tau_2$ for $i=1,2$, $\Gamma_{8}=\sigma_{3} \otimes \tau_3$, $\Gamma_{9}=\sigma_{0} \otimes \tau_3$ and $\Gamma_{0}=\sigma_{0} \otimes \tau_0$. Here ${\bm \sigma}$ and ${\bm \tau}$ represent orbital and spin degrees of freedom while writing the Hamiltonian in the basis $(c_{A\uparrow}, c_{A\downarrow}, c_{B\uparrow}, c_{B\downarrow})$.  The components $n_i$ are given by  
$n_0= 2 t_2 f({\bm k}) \cos \phi $, $n_1=-t_1 (1 + 2 h({\bm k})   )$ ,
$n_2= -2 t_1  \sin \frac{\sqrt{3}k_y}{2} \cos \frac{k_x}{2}$,
$n_3= M - 2 t_2 ~g({\bm k}) \sin \phi $,
$n_4= \frac{V_R}{\sqrt{3}} \sin \frac{\sqrt{3}k_y}{2} \cos \frac{k_x}{2} $,
$n_5= \frac{V_R}{\sqrt{3}} (h({\bm k}) - 1) $,
$n_6= - V_R \cos \frac{\sqrt{3}k_y}{2} \sin \frac{k_x}{2} $,
$n_7= V_R \sin \frac{\sqrt{3}k_y}{2} \sin \frac{k_x}{2}$,
$n_8= \frac{V_{\rm so}}{3} g({\bm k}) \cos \phi $, 
$n_9= \frac{V_{\rm so}}{3} f({\bm k}) \sin \phi$, with $f({\bm k})= 2 \cos \frac{\sqrt{3}k_y}{2} \cos \frac{k_x}{2} + \cos k_x$,
$g({\bm k})= 2 \cos \frac{\sqrt{3}k_y}{2} \sin \frac{k_x}{2} - \sin k_x $,
$h({\bm k})= \cos \frac{\sqrt{3}k_y}{2} \cos \frac{k_x}{2}$.  We note that for $V_R=V_{\rm so}=0$, the model (\ref{main:ham}) reduces to two copies of Haldane model (that breaks TRS, $ \mathcal T H({\bm k}) \mathcal T^{-1} \ne  H(-{\bm k} )$ with  $\mathcal T = (I \otimes \tau_2)i \mathcal K$, $\mathcal K$ being the complex conjugation) with spin up and down block. On the other hand, for $t_2=\phi=0$, it  reduces to Kane-Mele model (that preserves TRS).

{To understand the physical connection of the TRS breaking in the presence of intrinsic spin-orbit coupling, we investigate the following terms deeply:
$H_{A\uparrow A\uparrow}(\bm k)= n_0 + n_3 + n_8 + n_9=
( f(\bm k) \frac{V_{\rm so}}{3} - 2  g(\bm k)t_2 ) \sin \phi + ( g(\bm k) \frac{V_{\rm so}}{3} - 2 f(\bm k) t_2) \cos \phi  + M $ and 
$H_{A\downarrow A\downarrow}(\bm k)= n_0 + n_3 -n_8 -n_9 =
-( f(\bm k) \frac{V_{\rm so}}{3} + 2  g(\bm k)t_2 ) \sin \phi - ( g(\bm k) \frac{V_{\rm so}}{3} + 2 f(\bm k) t_2) \cos \phi  + M $.  We notice that in the absence of $V_{\rm so}$, $H_{A\uparrow A\uparrow}(\bm k)=H_{A\downarrow A\downarrow}(\bm k)$.  The 
similar line of argument is also applicable for $H_{B\uparrow B\uparrow}(\bm k)$ and $H_{B\downarrow B\downarrow}(\bm k)$.
In the semi-classical picture this refers to a situation  when the spin degrees of freedom are suppressed and only the orbital / charge degrees of freedom remain active just as the case for Haldane model. 
The band inversion conditions, estimated at Dirac points, take the same form irrespective of their spin components. However when the spin-orbit interaction is considered, we observe that it affects  the different spin components in the opposite way. This can be thought as a ${\bm k}$-dependent Zeeman field splitting between spin up and down components.  Thus we see a competition in energy scale due to orbital degrees of freedom and spin degrees of freedom. As a result, the band inversion condition of  both the spin component at the two Dirac points are no longer interdependent.} In the rest of the paper, we consider $t_1=1.0$ and $t_2=0.5$ without loss of generality. \\

\indent

 {Now it may be pertinent to digress a little and discuss about the
topological characterization of various phases. 
We may note that to describe two-dimensional electronic system involving explicit spin degrees of freedom, two topological invariants namely, $\text{Z}_2$ invariant \cite{kane-2005-2nd} and spin Chern number 
\cite{haldane06} were proposed in close succession. In the present case with 
TRS breaking magnetic flux, the spin Chern number continues to work while $\text{Z}_2$ invariant ceases \cite{prodan-2009}. 
We hence use the  spin Chern number ${\rm C}_{\uparrow}$ and ${\rm C}_{\downarrow}$
to classify different phases of the Hamiltonian as given in Eq.~(\ref{main:hamk}). In order to numerically compute the spin Chern number, 
one has to construct the projector $P(\bm k) = \Ket{V_1 (\bm k)}\Bra{V_1(\bm k) } + \Ket{V_2 (\bm k)}\Bra{V_2(\bm k) }$ 
with  $\Ket{V_1(\bm k)}$ and $\Ket{V_2(\bm k) }$ being the eigenvectors corresponding to two valence bands with energies $E_1(\bm k),~ E_2(\bm k)<0$. 
Now by diagonalizing the 4-dimensional projected spin operator $\tilde{S} (\bm k) = P(\bm k) (\sigma_0 \otimes \tau_3) P(\bm k)$, we can obtain four eigenvectors
$|\psi_{1,2,3,4}(\bm k) \rangle$  corresponding to four eigenvalues $\epsilon_{1,2,3,4}(\bm k)$ with $|\epsilon_1(\bm k)|=\epsilon_4(\bm k) \ne 0$, $|\epsilon_2(\bm k)|=\epsilon_3(\bm k) \simeq 0$ (within numerical accuracy) and $\epsilon_1(\bm k)<\epsilon_4(\bm k)$.
We further use the 4-component eigenvectors $|\psi_{1}(\bm k)\rangle$ and $|\psi_{4}(\bm k)\rangle$, corresponding to two non-zero eigenvalues,
to numerically compute ${\rm C}_{\uparrow}$ and ${\rm C}_{\downarrow}$, respectively. Here we  follow the Fukui's method in the ${\bm k}$-space to compute them \cite{spin-chern2}. Instead of constructing 4-dimensional projected spin operator $\tilde{S}(\bm k)$, using the property of projector operator $P(\bm k)$,
a 2-dimensional effective projected spin operator ${\mathcal S}_{ij} (\bm k)=\langle V_i(\bm k) | \sigma_0 \otimes \tau_3 | V_j (\bm k) \rangle$ with $i,j=1,2$,
can be alternatively used to compute the spin Chern numbers  \cite{sheng-2011,spin-chern-number}.}


To begin with, we show the phase diagram in $M-\phi$ plane  by  keeping $V_{\rm so}=1.0$ fixed as shown in Fig.~\ref{fig:modificationHaldane} (a) and (b) for $V_R=0$ and $0.5$, respectively.  As the modification over Haldane's phase diagram, we find that a finite $V_{\rm so}$ in Hamiltonian (\ref{main:ham}) results in two additional topological phases namely, QSHI [$(\rm{C}_{\uparrow},\rm{C}_{\downarrow})$ with {$ \rm{C}_{\uparrow} = -\rm{C}_{\downarrow} = \pm 1$}] and QASHI [$(\rm{C}_{\uparrow}= 0,\rm{C}_{\downarrow}= \pm 1)$ or $(\rm{C}_{\uparrow}= \pm 1,\rm{C}_{\downarrow}= 0)$]  phases.  The size of QAHI phases, characterized by spin Chern number $ (\rm{C}_{\uparrow},{C}_{\downarrow})$ with $\rm{C}_{\uparrow}=\rm{C}_{\downarrow}=\pm 1$, gets reduced as compared to QAHI phases in the  Haldane model; 
$(1,1)$ and $(-1,-1)$ phases are respectively encapsulated by QASHI phases $(0,1)$, $(1,0)$ and  $(0,-1)$, $(-1,0)$ from below and above. While the two adjacent QAHI phases are connected by QSHI phases $(1,-1)$ and $(-1,1)$. The color coded phase boundaries are assigned to the zeros of the respective energy gap equations \cite{phase_boundary}.  It is noteworthy  that  $\phi \rightarrow - \phi$, implies $\rm{C}_{\uparrow} \to - \rm{C}_{\downarrow}$ and 
$\rm{C}_{\downarrow} \to - \rm{C}_{\uparrow}$ for QASHI and QAHI phases. This correspondence holds also for QSHI phase that maps to itself.
The underlying reason could be the helical edge modes are time reversed partner of each other in QSHI phase.

Strikingly, we encounter an extended critical phase, denoted by assembly of black dots,   within which the spin gap vanishes identically as shown in Fig.~\ref{fig:modificationHaldane} (a) \cite{spin-gap}. This cap like critical phase can not be characterized by the spin Chern number. The vertical height (horizontal width) of the critical phase decreases (increases) with increasing $V_R$ (such that  $ V_R \leq V_{\rm so}$) while the 
size of QASHI phases reduces without qualitatively deforming their phase boundaries. The QASHI phases vanish and critical phase extends between $-\pi<\phi< \pi$
when $V_R > V_{\rm so}$ as   depicted in Fig. \ref{fig:modificationHaldane2} (a) and (b). In other words, 
the critical phases are bounded by violet phase boundaries from
outside for $V_R > V_{\rm so}$. 
This phase becomes the widest when $V_{\rm so}=0$   (see Fig. \ref{fig:modificationHaldane2} (a)). Upon introduction of $V_{\rm so}$, the size of QAHI phases reduces as well as critical phase becomes narrower (see Fig. \ref{fig:modificationHaldane2} (b)). Finally, when $V_{\rm so} \geq V_R$, QASHI phases start to appear 
near $\phi=0$ and $\pm \pi$. The violet phase boundaries expand with increasing $V_R$ and it fully extends $-\pi < \phi <\pi$ when $V_R \geq V_{\rm so}$. The gapless critical phase, otherwise bounded from outside, will now be bounded from inside by the violet phase boundaries as soon as  $V_{\rm so}$ exceeds $V_R$. The exact relation between $V_{\rm so}$ and $V_R$ can be found from the gap equation corresponding to the violet phase boundary \cite{phase_boundary}.

%


We now investigate the phase diagram in $V_R$-$M$ plane to elucidate the modification over Kane-Mele phases namely, QSHI phases as shown in Fig.~\ref{fig:modificationKaneMele} (a) and (b) for $\phi=0$ and $-\pi/4$, respectively. The distinctive feature is that finite $\phi$ is able to break the QSHI phase $(1,-1)$ into  QASHI phase $(1,0)$ and QAHI phase $(1,1)$, while NNN hopping $t_2$ alone does not affect the existing phase diagram. Notably, QAHI phase originates between the two lobes of QASHI phase. We note that  TRS  breaking uniform exchange field can lead to QAHI  phases \cite{sheng-2011}. The staggered magnetic flux $\phi$ associated with NNN hopping $t_2$ and spin dependent hopping $V_{\rm so}$ acts as a key ingredient to generate all the above phases simultaneously. It is to be noted that QASHI phases appear when $V_R< V_{\rm so}$. The color coded phase boundaries indicates that the identical QASHI phases for positive and negative $M$  are bounded by same gap equations.

Below, we emphasize a few essential conclusions from these phase diagrams. The QAHI (QSHI) lobes of Haldane (Kane-Mele) model dismantle into a variety of phases in the  presence of $V_R$ and $V_{\rm so}$ ($t_2$ and $\phi$) giving rise to multicritical points where multiple topological phase boundaries coalesce. Across a phase boundary, separating two topological phases,
$|\Delta \rm{C}_{\uparrow} + \Delta \rm{C}_{\downarrow}|$ can only become unity where $\Delta \rm{C}_{\uparrow}$ ($\Delta \rm{C}_{\downarrow}$) measures the difference in $\rm{C}_{\uparrow}$ ($\rm{C}_{\downarrow}$) among the two adjacent topological phases separated by a phase boundary. This situation no longer holds generically 
when we encounter a multicritical point.  The most important finding of our work is the emergence of QASHI phase where only one spin component is topologically protected leaving the other to be trivially gapped out. Even though, this type of phase has been found in magnetically doped QSHI material \cite{Liu-2008,li2013,qiao10}, ours is the first tight binding model hosting these phases naturally, to the best of our knowledge.

\noindent

\noindent


We  now describe the edge state in the zig-zag edge ribbon geometry (periodic in $x$ direction and finite in $y$ direction) and its connection to the bulk invariant as a probe to understand the
bulk-boundary correspondence \cite{nag19}. In Fig. \ref{fig:edgeStateVsoGTVr} (a), (b), (c) and (d), we depict the edge modes for $(1,0)$, $(-1,0)$, $(0,1)$ and $(0,-1)$, respectively. To this end we generalize the  bulk-boundary correspondence for QAHI and QSHI  \cite{haldane-1988, kane-2005-2nd} to QASHI: $\rm{C}_{\sigma}= N^{\sigma}_{\rm RM}- N^{\sigma}_{\rm LM}$, where  $N^{\sigma}_{\rm RM}$ ($N^{\sigma}_{\rm LM}$) represents the number of right (left) moving edge mode for spin $\sigma=\uparrow,\downarrow$.  We note that edge modes are not completely spin polarized as far as their numerical calculations are concerned. 
We assign an edge state to be spin up (down) if it is maximally populated by spin up (down) states. 
Turning to the helical edge states in QSHI phase, as shown in Fig.~\ref{fig:edgeStateVsoGTVr}(e) and (f), the spin dependent chiral motion is clearly captured where up (down) spin traverses in a clockwise (anti-clockwise) manner along the edges of the system. 
This results in two types of QSHI phases with  spin Chern number $(1,-1)$ and $(-1,1)$ depending on the chirality of the spin-polarized edge state.
Finally, we show the chiral edge states of $(1,1)$ and $(-1,-1)$ QAHI phases respectively in Fig.~\ref{fig:edgeStateVsoGTVr} (g) and (h). In this case, 
both the spin up and down edge states share same chirality while  
traversing along the boundaries of the system. Therefore, using the bulk-boundary correspondence, we can successfully explain 
that the spin dependent edge states in different phases are related by the spin Chern number of the underlying phases.

Another very intriguing fact that we notice is  except for $\phi=\pm\pi/2$, where QAHI phases host zero energy chiral edge states as shown in Fig.~\ref{fig:edgeStateVsoGTVr} (g) and (h), all the other values of 
$\phi\ne \pm \pi/2$ support finite energy edge states if there exist  a topological phase. The edge modes do not show any avoided level crossing structure that are observed for  QSHI in presence of magnetically doping  and exchange field 
\cite{sheng-2011,Liu-2008,li2013}. Therefore, staggered flux induced topological phases are intrinsically different from the above cases even though the TRS is broken in both the situations. An effective TRS emerges implying $E(\pi-k, \phi)= E(\pi+ { k}, - \phi)$ in our case. 
Even more surprisingly, edge modes are further protected by 
a composite anti-unitary symmetry ensuring $E(\pi -{ k},\pi -\phi)=-E(\pi+{k},\pi + \phi)$.  Thus the twin effect of these anti-unitary symmetries allows one the mapping $\rm C_{\uparrow} \to - \rm C_{\downarrow}$ and $\rm C_{\downarrow} \to - \rm C_{\uparrow}$ under $\phi \to -\phi$. 
This further guarantees  the existence of zero energy chiral edge modes for any topological phase obtained at $\phi= \pm \pi/2$.

Having extensively explored the lattice model, we now make resort to the low energy model for better understanding behind the emergence of different phases. Expanding around the Dirac points $\alpha=\pm 1$, we obtain $n_1= - \alpha \sqrt{3} t_1 k_x/2$, $n_2=\sqrt{3} t_1 k_y/2$, $n_3= M + \alpha 3 \sqrt{3} t_2 \sin \phi \theta({\bf k}) $, $n_4=-V_R k_y/4$, $n_5=- V_R/\sqrt{3} +\alpha V_R k_x/4$, $n_6=\alpha V_R/\sqrt{3} + V_R k_x/4$, $n_7= -\alpha 3 V_R k_y/4$, $n_8=-\alpha (\sqrt{3}/2) V_{\rm so} \cos \phi \theta(\bf k)$, $n_9= - (V_{\rm so}/2) \sin \phi \theta({\bf k})$ with $\theta({\bf k})= (1- |k|^2/4)$. At the Dirac points, the eigen-energies take the following form:  $E_{1,4}= (w_1 + w_4 \pm  \lambda^{14}_1)/2$, $E_{2,3}= (w_2 + w_3 \pm \lambda^{23}_2)/2$  with $w_1=n_3 + n_8 + n_9$, $w_2=n_3-n_8-n_9$, $w_3=-n_3 -n_8 + n_9$, $w_4=-n_3 +n_8 - n_9$, $r_1=-n_5 - n_6$ and $r_2=n_6-n_5, \lambda^{jk}_i=\sqrt{4 r^2_i + (w_j-w_k)^2}$. Let us now start with a simple case $V_R=0$ leading to the energy gap for spin up $\Delta E^{\uparrow}_{AB}=w_1-w_3$, and spin down  $\Delta E^{\downarrow}_{AB}= w_2 -w_4$. In this case, the low energy model closely follows the Bernevig-Hughes-Zhang model for HgTe quantum Well \cite{bhz1}  enabling us investigate different phases in the similar spirit.  
A topological phase is ensured by  opposite signs of the gap at two Dirac points ${\bm k}_1$ and ${\bm k}_2$:
$\Delta E^{\uparrow(\downarrow)}_{AB} ({\bm k}_1) \Delta E^{\uparrow(\downarrow)}_{AB} ({\bm k}_2) <0$.
The different combination of the above product can in principle determine various topological phases.

For the QASHI phases with $(\rm{C}_{\uparrow},0)$ [$(0,\rm{C}_{\downarrow})$], one can find spin up [down] sector is only topologically  gapped  out leaving other  spin sector to be trivial. For the QSHI phase with $(\rm{C}_{\uparrow}=\pm 1,\rm{C}_{\downarrow}=\mp 1)$, one can find different combinations of topological gap in both the spin sectors. In the case of QAHI phase
with $(\rm{C}_{\uparrow}=\pm 1,\rm{C}_{\downarrow}=\pm 1)$, the same combination of topological gap occur in both the spin sectors.  Denoting $x_{\zeta,\xi}=M + \zeta 3 \sqrt{3} t_2 \sin \phi + \xi (V_{\rm so}/2) \sin \phi$, we find that for topological spin up channel with $\rm{C}_{\uparrow} \ne 0$, 
$x_{+-}x_{-+}<0$; on the other hand, for  topological spin down channel with $\rm{C}_{\downarrow} \ne 0$, 
$x_{++}x_{--} < 0$. The phase boundaries across which $\rm{C}_{\uparrow}$ ($\rm{C}_{\downarrow}$) changes are  obtained by solving for $M$ from $x_{\pm\mp} $ $(x_{\pm\pm})$=0. This further explains the observation that $\rm{C}_{\uparrow}$ and $\rm{C}_{\downarrow}$  can only jump by unity across a  phase boundary separating two different topological phases.  However, there exist multicritical points in the phase diagram where more than two phases converge including non-topological phases. At these points, spin Chern number can jump by more than unity. Without Rasbha interaction $V_R =0$, one can observe QAHI, QSHI and QASHI phases in various parameter regimes as shown in  Fig.~\ref{fig:modificationHaldane} and Fig.~\ref{fig:modificationKaneMele}, can be
 explained by the above low energy analysis.

Now, we extend our analysis for finite $V_R \ne 0$ using the low energy formulation. We note that phase boundaries are modified
without altering the topological nature of the phases
in presence of $V_R$ provided  $V_{\rm so} \ne 0$ and $\phi \ne 0$. This suggests that phases, present  in the absence of $V_R$,  are adiabatically connected while Rashba interaction is turned on. The principle for a phase being topological remains unaltered, however, their explicit forms are modified. Denoting $y_{\eta}= \eta \sqrt{3} (V_{\rm so}/2) \cos \phi$ and $z_{\zeta,\xi}=\sqrt{ 4 V_R^2/3 + x^2_{\zeta,\xi}}$, we find 
$\Delta E^{\uparrow \downarrow}_{AB}({\bm k}_1)= \pm  y_{-} + |x_{+,-}| {\rm sgn}(x_{+,-}) + z_{+,+} $, $\Delta E^{\uparrow \downarrow}_{AB}({\bm k}_2)= \pm y_{+} + |x_{-,+}| {\rm sgn}(x_{-,+}) + z_{-,-} $. For spin up (down) sector to be topological, the following condition needs to be satisfied 
$[y_{-} + |x_{+,-}| {\rm sgn}(x_{+,-}) + z_{+,+}][y_{+} + |x_{-,+}| {\rm sgn}(x_{-,+}) + z_{-,-}] < 0$ ($[y_{+} + |x_{+,-}| {\rm sgn}(x_{+,-}) + z_{+,+}][y_{-} + |x_{-,+} |{\rm sgn}(x_{-,+}) + z_{-,-}] < 0$). In order to obtain physical phase boundaries $z_{\zeta,\xi}^2$ has to be positive that yields the modification of phase boundaries in presence of $V_R$
\cite{phase_boundary}.
It is important to note that in addition to the $V_R=0$ case, the relative strength between  $V_{\rm so} \sin \phi$ and $\sqrt{3}V_R/2$ terms also play an important role in determining the phase boundaries. For example, the phase boundary is substantially modified with new topological phase once $\phi$ becomes non-zero (i.e., TRS is broken) as shown in Fig.~\ref{fig:modificationKaneMele} (a) and (b). 
Relying on the structure of topological gap at Dirac points, one can define the spin Chern number in an effective manner as follows 
\begin{eqnarray}
 \rm{C}_{\sigma}=&&\frac{1}{2} \Bigg[{\rm sgn} \bigg(\Delta E^{\sigma}_{AB}({\bm k}_2)\bigg) - {\rm sgn} \bigg(\Delta E^{\sigma}_{AB}({\bm k}_1)\bigg) \Bigg],
 \label{sc_number}
\end{eqnarray}
 with $\sigma=\uparrow,\downarrow$. Now we elaborate the role of  the spin gap defined as $\Delta \mathcal{E}^{\uparrow \downarrow}_{A}= E^{\uparrow}_A - E^{\downarrow}_A $, and $\Delta \mathcal{E}^{\uparrow \downarrow}_{B}= E^{\uparrow}_B - E^{\downarrow}_B $ at two Dirac points in examining underlying stability for topological transitions. It is indeed necessary to have finite spin gap 
$\Delta \mathcal{E}^{\uparrow \downarrow}_{A}, \Delta \mathcal{E}^{\uparrow \downarrow}_{B} \ne 0$, in order to characterize a phase with $(\rm{C}_{\uparrow},\rm{C}_{\downarrow})$ \cite{spin-gap}. Therefore, the robustness and stability of the topological invariant is determined by the finiteness of the spin gap. 
Interestingly, our numerical calculation with lattice model suggests that spin gap can vanish at any arbitrary point inside the momentum BZ for $\phi \ne 0$. This is in contrast to the energy gap, obtained from  the lattice model, that only vanishes at Dirac points.

In conclusion, we consider the TRS broken Kane-Mele model merged with Haldane model where intrinsic SOC is coupled with staggered magnetic flux to investigate the fate of QSHI phases. We remarkably find new topological phases namely, QASHI phase and extended critical region in addition to the QAHI and QSHI phase while studying the spin Chern number $(\rm{C}_{\uparrow},\rm{C}_{\downarrow})$. The QASHI phase, characterized by $(0,\rm{C}_{\downarrow})$ and $(\rm{C}_{\uparrow},0)$, supports spin-selective transport where 
one spin channel is topologically gapped out leaving the other component trivially gapped. The other two topological phases namely, QSHI and QAHI phases exhibit spin-polarized and spin-neutral edge transport in accordance with spin Chern number. 
{In short, superimposing Haldane model with Kane-Mele model,
we can successfully unify all possible types of QH phases for two dimensional non-interacting system in a single phase diagram.}
The topological phases in this model are preserved by
an effective TRS and a composite particle-hole symmetry.
We show that the findings from the lattice model can be understood from 
low energy model around the Dirac point. We also provide an effective description of spin Chern number, based on the low energy model, that 
corroborates with the lattice calculation. Surprisingly, the band gap turns out to be decisive in the topological characterization while stability and robustness  is determined by the finiteness of the spin gap.  In terms of the future applications, our study can become useful in exploring the spin entanglement in different phase ~\cite{soc_entanglement,2d_material} and disorder induced Anderson QSHI phases 
\cite{soc_anderson}.

{
Before ending we shall discuss the connection of work to recent experimental advancements.} 
 We note that in optical lattice platform SOC
is theoretically proposed \cite{soc_theory1,soc_theory2,soc_theory3,soc_theory4,soc_theory5} and experimentally realized \cite{soc_exp1,soc_exp3,soc_exp4,soc_exp5}. {In particular, the method of producing complex next-nearest neighbour hopping, that has been employed earlier  to realize the Haldane model, can be utilized further to the case of spin dependent hopping with complex amplitudes \cite{exp1}. The suggested procedure involves the application of spin-dependent force that is caused by oscillating magnetic field gradient. By adjusting the oscillating field and the mirror position together the complex phase factor, appearing for the opposite spin component, may be engineered.  As far as the real materials are concerned there have been several proposals and experimental realizations of QAH effect which include magnetic-ion-doped HgTe quantum well \cite{Liu-2008}, topological insulator surfaces \cite{yu-2010},  transition metal oxides \cite{xiao-2011,cook-2014} and engineered graphene \cite{qiao10,zhang-2012}. 
The Haldane like complex next nearest hopping amplitude was found in a series of Fe-based honeycomb ferromagnetic insulators, $\rm{AFe(PO_4)_2(A=Ba, Cs, K, La)}$ which possess Chern bands \cite{hskim-2017}.
The spin dependent hopping can be engineered in the materials such as,  $\rm{AFe(PO_4)_2}$,  graphene with magnetic impurity \cite{qiao10,qioa-2014}, magnetically ordered two dimensional materials \cite{chang-2019,ramon-2020},
transition metal oxide hetero-structures~\cite{xiao-2011},  Skyrmion lattice \cite{muhlbauer-2011,kurumaji-2019,choi-2021,gilbert-2015} and magnetic insulators $\rm{MnTe, MnSe}$ that 
could then serve as the potential candidates to realize our model in principle. 
On the other hand, 
the materials with intrinsic spin-orbit coupling that host QAH effect~\cite{kim16,garrity-2013} may be promising candidates if the spin-orbit interaction is controlled by using the methods outlined before. Moreover the external pressure could be a useful way to control the Rashba spin interaction~\cite{huang-2020} whereas the intrinsic spin orbit interaction can be manipulated by suitable doping and other interactions \cite{kandemir-2013,luis-2015,tobias-2017}.  Though the quantitative prescription  of such schemes is beyond the scope of present study and will be presented elsewhere, we expect that with the state of the art experiments in near future such  a controlled spin-dependent hopping can be realized. Apart from the possible experimental realizations, our work has significant amount of technological relevance in the context of  spintronics within modern electronics, spin field effect transistors  and magnetic field sensors of hard disk drives
\cite{spintronics1,spintronics2,spintronics3}.} 

{\it Acknowledgement}: S.S would like to thank SAMKHYA: High Performance Computing Facility provided by Institute of Physics, Bhubaneswar. T.N thanks Bitan Roy for useful discussions. S.M thanks A. K. Nandy for useful discussions.


\begin{thebibliography}{100}
\bibitem{spin-gap} The spin gap is calculated from the spin projected Hamiltonian
$H_{ij}(\bm k)=\langle \psi_i({\bm k}) | \sigma_0 \tau_3| \psi_i({\bm k}) \rangle $  where 
$|\psi_i({\bm k})\rangle$ denote the valence band wave-functions $i=1,2$.
\bibitem{phase_boundary}
Using the low energy expansion around the Dirac points($\alpha=\pm$), we  obtain the following  gap equations from four eigen-energies:
$ \delta E_{1}^\alpha = -\alpha M + \eta_{1,-} - \Delta_1 $, 
$\delta E_{2}^\alpha = -\alpha M + \eta_{1,-} + \Delta_1$, 
$\delta E_{3}^\alpha = 2\Delta_1$, 
 $\delta E_{4}^\alpha  = -2 \alpha M + \eta_{0,-}$, 
$ \delta E_{5}^\alpha = -\alpha M + \eta_{-1,-} + \Delta_1 $,
$ \delta E_{6}^{\alpha} = -\alpha M + \eta_{-1,-} - \Delta_1$ with $\eta_{a,\pm}=\sqrt{3} aV_{\rm so} \cos \phi + \frac{1}{2}(V_{\rm so} \pm 6 \sqrt{3} t_2) \sin \phi$,   $  \Delta_1 =  \left( M^2+ 3V^2_R + (27t^2_2+ 3 \sqrt{3} t_2 V_{\rm so} + \frac{V^2_{\rm so}}{4}) \sin^2 \phi - 2 \alpha M \eta_{0,+} \right)$.
\end{thebibliography}
\end{document}